\documentclass[11pt,a4paper]{article}

\usepackage{jheppub}
\usepackage{color}

\def\dfrac#1#2{\displaystyle\frac{#1}{#2}}

\newcommand{\eq}[1]{Eq.~(\ref{#1})}

\newcommand{\pslash}{p\kern-1ex /}
\newcommand{\qslash}{q\kern-1ex /}
\newcommand{\lslash}{l\kern-1ex /}
\newcommand{\sslash}{s\kern-1ex /}
\newcommand{\kaslash}{k_a\kern-2ex /}
\newcommand{\kbslash}{k_b\kern-2ex /}
\newcommand{\Dslash}{{\cal D}\kern-1.5ex /}

\newcommand{\beqa}{\begin{eqnarray}}
\newcommand{\eeqa}{\end{eqnarray}}

\begin{document}
\title{Do we know how to define energy in general relativity ?}
\author[a]{Sinya AOKI}
\affiliation[a]{Center for Gravitational Physics and Quantum Information, Yukawa Institute for Theoretical Physics, Kyoto University, Kitashirakawa Oiwakecho, Sakyo-ku, Kyoto 606-8502, Japan}
\abstract{This essay is dedicated to Prof. K.K. Phua on the occasion of his 80th birthday.
While the contents of this essay are based on our recent papers\cite{Aoki:2020prb,Aoki:2020nzm} published in the International Journal of Modern Physics A,
I have added many personal opinions, so that I am solely responsible for all statements in this essay.
  }
\preprint{YITP-22-88}
\maketitle

\section{Introduction}
You may  think that an answer to the question in the title of this essay must be ``Yes", since  
a concept of energy is essentially important in physics so that it can be well understood 
in general relativity, which has now matured more than 100 year after Einstein's first proposal in 1915\cite{Einstein:1916}.
As I will show in this essay, however, the correct answer is  probably ``No"  or  more precisely ``A concept of energy is not well understood
in general relativity." .

Let me start my discussion with Einstein equation in general relativity, which reads
\beqa
R^{\mu\nu} -{1\over 2} g^{\mu\nu} R = 8\pi G T^{\mu\nu}
\label{eq:EE}
\eeqa
where $T^{\mu\nu}$ on the right-hand side is an energy momentum tensor (EMT) of matter, while the Ricci tensor $R^{\mu\nu}$, the Ricci scalar $R$ and the metric $g^{\mu\nu}$ on the left-hand side describes how a spacetime is curved,
and  $G$ is the Newton constant, which  controls a coupling strength of matter with the spacetime. 
The Newton constant is very small as $G =6.7\times 10^{-39}$ (GeV)$^{-2}$ in Natural unit ($c=\hbar =1$).
The Einstein equation tells us how existing matter determines the structure of curved spacetime.

Einstein derived this equation from two assumptions ---- the invariance under the general coordinate transformation (general covariance) and the equivalence principle. 
The former means that  the law of gravity in physics  does not depend on a choice of a coordinate system that describes spacetime,
while the latter says that  a gravitational force can be  locally removed away  by the coordinate transformation. 
 
\section{Dilemma: Conserved energy or coordinate independence ?}
A covariant derivative applied to the left-hand side of the Einstein equation vanishes identically. 
This is a consequence of the {\it Bianchi identity},  which implies the covariant conservation of the EMT as
\beqa
\nabla_\mu T^{\mu\nu} := \partial_\mu T^{\mu\nu} + \Gamma^\mu_{\mu\alpha} T^{\alpha\nu} +\Gamma^\nu_{\mu\alpha}T^{\mu\alpha}=0,
\label{eq:convEMT}
\eeqa
as a generalization of that in flat spacetime, $\partial_\mu T^{\mu\nu}=0$, where
$\Gamma^{\nu}_{\mu\alpha}$ is called a connection or a Christoffel symbol.
Generally, eq.~\eq{eq:convEMT} does not identically holds but is satisfied only after equations of motions for matter are used. 
While a conservation of energy in flat spacetime is derived from $\partial_\mu T^{\mu\nu}=0$, 
eq.~\eq{eq:convEMT} does not provide a conserved energy in general, due to extra second and third contributions in the covariant derivative.
Then, what is conserved energy in general relativity ?

One of the textbook answers was proposed by Einstein himself.
He modified the EMT as $\tilde T^{\mu\nu} := T^{\mu\nu} + t^{\mu\nu}$, which satisfies
\beqa
\partial_\mu (\sqrt{-g} \widetilde{T}^{\mu\nu}) &=& 0, 
\label{eq:pseudo-T}
\eeqa 
where $g$ is a determinant of $g_{\mu\nu}$  and $\sqrt{-g}$ is a spacetime volume density.
Thus a conserved energy may be defined by
\beqa
E = \int d^3x\, \sqrt{-g} \widetilde{T}_{00}  \quad \mbox{or} \quad   E = -\int d^3x\, \sqrt{-g} \widetilde{T}^{0}{}_0,
\eeqa
where an integral measure $\sqrt{-g}\, d^3x$ is covariant under general coordinate transformations.
Unfortunately there exists no theoretical preference between $\widetilde{T}_{00}$ and $\widetilde{T}^{0}{}_0$ in this approach.
A problem of Einsteins's such approach is that $t^{\mu\nu}$ is not a tensor under general coordinate transformations, as evident from that 
\eq{eq:pseudo-T} is not a covariant equation. 
Thus $\widetilde{T}^{\mu\nu}$ does not satisfy one of the two fundamental assumptions in general relativity,  the general covariance, 
and $\widetilde{T}^{\mu\nu}$ is called Einstein's pseudo-tensor due to this violation. 

There is a dilemma about the choice between non-conservation of energy  or the violation of the general covariance. In my opinion Einstein made the worse choice, but after some debates on its limitations and problems, the pseudo-tensor approach has been gradually accepted for the definition of energy in general relativity, partly based on arguments that $t^{\mu\nu}$ represents the energy of gravitational fields, which hence must be non-covariant and coordinate dependent
since the equivalence principle always leads to gravitational fields that vanish by a coordinate transformation.
I feel that this view, found frequently in some textbooks, is not scientifically convincing to guarantee the correctness of the pseudo-tensor, since
we may invert the logic that we disfavor the pseudo-tensor since we can always make $t_{\mu\nu}$ zero locally.

Later there appeared conserved and covariant definitions, called quasi-local energies,
which give a total energy only as surface integrals without knowing a local energy density,
and include the well-known ADM energy\cite{Arnowitt:1962hi}.
Quasi-local energies, however, cannot solve our dilemma,  since their conservation is merely an identity as a consequence of general coordinate transformations,
known as the Noether's 2nd theorem\cite{Noether:1918zz}, and this flaw is also true for the pseudo-tensor\cite{Aoki:2022gez}. 
Under such unsatisfactory circumstance for a definition of energy in general relativity, 
we have recently proposed an alternative, which will be explain in this essay.

\section{Alternative definition of energy in general relativity}  
In Ref.~\cite{Aoki:2020prb},  
we consider a case that a metric allows a time-like or stationary
Killing vector satisfying $\nabla_\mu \xi_\nu + \nabla_\nu \xi_\mu =0$.
For example, if a metric is independent of a time coordinate $x^0$, the stationary Killing vector is given by $\xi^\mu=\xi^\mu_0:=-\delta^\mu_0$,
where a minus sign is chosen for energy, given later,  positive.
In this case, 
we have proposed an alternative definition of energy in general relativity as
\beqa
E_{\rm ours} = \int_\Sigma [d^3x]_\mu\, \sqrt{-g} T^{\mu}{}_\nu \xi^\nu,
\label{eq:energy}
\eeqa 
that is conserved since $\partial_\mu (\sqrt{-g} T^{\mu}{}_\nu \xi^\nu) =  \sqrt{-g} \nabla_\mu (  T^{\mu}{}_\nu \xi^\nu ) =  \sqrt{-g} 
T^{\mu\nu} (\nabla_\mu\xi_\nu +\nabla_\nu\xi_\mu)/2 =0$,
where $\Sigma$ is a space-like entire hyper-surface with a volume factor  $ [d^3x]_\mu $.
Since  \eq{eq:energy} is  manifestly invariant under the general coordinate transformation, 
we confidently say ``Yes" to the question in the title of this essay in special cases while avoiding the dilemma.   
Unlike the pseudo-tensor and quasi-local energies, 
 \eq{eq:energy} passes a minimum test for an energy in general relativity to satisfy 
 that it leads to the standard energy in flat spacetime  in the Newton constant $G\to 0$ limit as
\beqa
\lim_{G\to 0} E_{\rm ours} = -\int_{x^0={\rm fix}} d^3x\, T^0{}_0 =\int_{x^0={\rm fix}} d^3x\, T_{00}.
\eeqa

While \eq{eq:energy}  or similar ones  can be found in some textbooks, there were no application of the definition before
we wrote a paper\cite{Aoki:2020prb}, where several examples were considered.
As a simple example, an energy of the Schwarzschild blackhole becomes
\beqa
E_{\rm ours}^{BH} = \int d^3 x\, \sqrt{-g} T^0{}_\nu \xi_0^\nu =M, 
\label{eq:energyBN}
\eeqa
where the EMT of the blackhole is given as
\beqa
T^0{}_0 = -{1\over 4\pi} {\partial_r (M\theta(r) )\over r^2} =  -{1\over 4\pi} { M\delta(r) \over r^2},
\label{eq:EMT_BH}
\eeqa
which shows that the energy is concentrated at the origin producing singularity.
Thus, unlike a  folklore that the Schwarzschild blackhole is a vacuum solution to the Einstein equation \eqref{eq:EE},
it has the non-zero EMT at the origin so that
the Schwarzschild blackhole is indeed NOT a vacuum solution. 
Since only a local property near $r=0$ is relevant for this derivation,
a mechanism for non-zero EMT at $r=0$ is identical  to that 
which produces a $\delta$ function singularity for the point charge in the Maxwell equation.
As I have been rather surprised, however,  many peoples including those who are though to have the highest intelligence in the world claim or believe that
the Schwarzschild blackhole is a vacuum solution to the Einstein equation, even though physically a gravitational collapse of matter 
leads to a black holes as a final 
state.\footnote{It is often claimed that singularities of black holes in general relativity
cannot be described by distributions such as the $\delta$ function  due to nonlinearities of the theory. 
As shown in \eqref{eq:EMT_BH}, however, the singularity of the Schwarzschild blackhole indeed leads to  $\delta$ function singularity in the EMT without having products of distributions. Mathematically this means that the Einstein equation \eqref{eq:EE}  should be understood in a distributional sense\cite{Balasin:1993fn}.
A non-vanishing EMT has been calculated also for the Kerr blackhole\cite{Balasin:1993kf}.  }
I suspect that this is a reason why the formula \eqref{eq:energy} has not been used so for, though the definition has been known for a long time:
 If the Schwarzschild blackhole were a vacuum solution, 
 the definition \eqref{eq:energy}  would give zero,  an unacceptable result for  blackhole energy.

\section{Energy of a neutron star} 
In the case of the  Schwarzschild blackhole, 
there appears no  difference between the energy given in \eqref{eq:energy}
and the pseudo-tensor/quasi-local energy\cite{Aoki:2022gez,Aoki:2020prb}. 
We here consider energy of a compact star such as a neutron star,
which reveals a difference between the two\cite{Aoki:2020prb}.

Energy defined by \eqref{eq:energy} for a static and spherically symmetric compact star becomes
\beqa
E_{\rm our} &:=& -\int d^3x\, \sqrt{-g} T^0{}_0 = 4\pi \int_0^R dr\, \sqrt{-g_{00}(r) g_{rr}(r)} r^2 \rho(r),
\eeqa   
where $R$ is a radius of the compact star, $g_{00}(r)$ and $g_{rr}(r)$ are time and radial components of a diagonal metric, respectively,
and $\rho(r)$ is a matter density related to the EMT as $T^0{}_0 =-\rho(r)$.
On the other hand, the ADM energy, called Misner-Sharp mass, is given as
\beqa
E_{\rm ADM} = 4\pi  \int_0^R dr\,  r^2 \rho(r),
\eeqa 
which does not have the $\sqrt{-g}$ factor, and thus is not a covariant. Note that, since $g_{00}(r) g_{rr}(r) =-1$ for the Schwarzschild blackhole,
 both definitions agree.

Let us compare these two energies in the Newtonian limit that $G$ is small and ${P(r)\ll \rho(r)}$ where $P(r)$ is a pressure of the matter.
In this limit we have
\beqa
E_{\rm our} - E_{\rm ADM} &=&  {G\over 2} \int d^3x \rho_0(\vert\vec x\vert) \phi(\vec x) + O(G^2, G \omega), \quad
\omega :=\displaystyle \max_r {P(r)\over \rho(r)},
\eeqa
where $\rho_0(\vert \vec x\vert)$ is the matter density at $\vec x$ in this limit and
$\phi(\vec x)$ is the gravitational potential at the point. 
On the other hands, $E_{\rm ADM}$ in the Newtonian limit  becomes
\beqa
E_{\rm ADM} = \int d^3x\, \rho_0(\vert\vec x\vert) + {G\over 2} \int d^3x \rho_0(\vert\vec x\vert) \phi(\vec x) + O(G^2, G \omega) .
\eeqa 
Thus $E_{\rm our}$ represents the energy of matter in the presence of the background gravitational potential $\phi(\vec x)$, while
$E_{\rm ADM}$ is the energy of the self-interacting matters,\footnote{I would like to thank Prof. Masaru Shibata for pointing out this difference to me.}
where
\beqa
\phi(\vec x) =-  \int d^3y {\rho_0(\vert\vec y\vert) \over \vert \vec x-\vec y\vert} .
\eeqa
Therefore both seem reasonable for different situations.\footnote{Note however that this conclusion is different from the one in \cite{Aoki:2020prb}.
Thus I am solely responsible for the content in this section.}
It will be interesting to understand the meaning of the difference.

\section{Generalization}
The energy given in \eqref{eq:energy} is not the final answer, since a metric usually does not have a Killing vector $\xi_0^\nu$ in general.
Then, how can we define energy for a general metric in the absence of the Killing vector ?

While $\xi_0^\mu$ is no more a Killing vector for a general metric, 
the energy $E[\xi_0]$ in \eqref{eq:energy} is still conserved if the metric satisfies\cite{Aoki:2020nzm}, 
\beqa
 T^\mu{}_\nu \nabla_\mu\xi_0^\nu = - T^\mu{}_\nu \Gamma^\nu_{\mu 0} = 0,
\eeqa
where summations over $\mu,\nu$ are implicitly assumed.
The conservation of $E[\xi_0]$ is easily confirmed, 
since 
\beqa
\partial_\mu \left (\sqrt{-g} T^{\mu}{}_\nu \xi_0^\nu \right) = \sqrt{-g} \nabla_\mu ( T^{\mu}{}_\nu \xi_0^\nu ) =
 \sqrt{-g}  T^{\mu}{}_\nu  \nabla_\mu\ \xi_0^\nu =0, 
\eeqa
where we use $\nabla_\mu T^\mu{}_\nu = 0$ and a property of the covariant derivative $\sqrt{-g} \nabla_\mu J^\mu =\partial_\mu (\sqrt{-g} J^\mu )$ 
for an arbitrary vector $J^\mu$.
For example, the energy is indeed conserved during some types of gravitational collapses\cite{Aoki:2020nzm}.

For more general cases, however, $E[\xi_0]$ is no more conserved.
A simple example is a model of a homogeneous and isotropic expanding universe with the Friedmann-Lema\^itre-Robertson-Walker (FLRW) 
metric, given by
\beqa
ds^2 = -(dx^0)^2 + a^2(x^0)  d\vec x^2,
\eeqa
where $a(x^0)$ is a scale factor. 
The EMT is given by the perfect fluid as
\beqa
T^0{}_0 = -\rho(x^0), \quad T^i{}_j (x^0)=P(x^0) \delta^i_j, \quad T^0{}_j =T^i{}_0 =0,
\eeqa
whose conservation $\nabla_\mu T^\mu{}_\nu =0$ implies 
\beqa
\dot \rho +3(\rho+P) {\dot a\over a} =0, 
\eeqa
where $\dot f$ means a derivative of $f$ with respect to a time coordinate $x^0$.

The energy \eqref{eq:energy} is easily calculated as
\beqa
E := E[\xi_0] = -\int d^3x \sqrt{-g} T^0{}_0 = V_0 a^3\rho, 
\eeqa
where $V_0:= \int d^3 x$ is a constant space volume without the scale factor $a(x^0)$.
This energy is indeed not conserved as
\beqa
\dot E = V_0 ( \dot \rho a^3 +3 \rho \dot a a^2 ) = -3 {\dot a\over a} {P\over \rho} E\not= 0,
\eeqa
as long as $P\not=0$ and $\dot a\not=0$.

In Ref.~\cite{Aoki:2020nzm}, we have proposed a method to define a conserved quantity as a generalization of the energy,
which is given by
\beqa
S:= E[\beta \xi_0] =   -\int d^3x \sqrt{-g} T^0{}_0 \beta = \beta E, 
\eeqa
where $\zeta^\mu(x^0):=\beta(x^0)\xi_0^\mu$ should satisfy
\beqa
 T^\mu{}_\nu \nabla_\mu\zeta^\nu \left( =\rho\dot\beta -3P {\dot a\over a}\beta \right) = 0.
 \label{eq:beta}
\eeqa
Then it is easy to see that $S$ is conserved as
\beqa
\dot S = \beta \dot E + \dot \beta E = - 3 {\dot a\over a} {P\over \rho} \beta E + 3 {\dot a\over a} {P\over \rho} \beta E = 0.
\eeqa

What is the physical meaning of $S$ ?
Using \eqref{eq:beta},   we rewrite the above conservation equation as
\beqa
{d S\over dx^0} = {d E\over dx^0}\beta + E{d \beta\over dx^0} =\left( {d E\over dx^0} + P {d V\over dx^0}\right)\beta,
\label{eq:1st}
\eeqa
where $V(x^0) := V_0 a^3(x^0)$ is a time-dependent space volume during the expansion.
Since  \eqref{eq:1st} is similar to the first law of thermodynamics, 
\beqa
T dS = dE + P d V,
\eeqa
we interpret $S$ as the total {\it entropy} of the universe and $\beta=\dfrac{1}{T}$ as an inverse temperature.
Thus, although the total energy is not conserved, the total entropy is conserved during the expansion.
In addition,  $\beta(x^0)$ gives the time-dependent inverse temperature of the universe, which increases (equivalently
the temperature decreases) as the universe expands, since
\beqa
{\dot \beta\over \beta} = 3{P\over \rho} {\dot a\over a} > 0.
\eeqa

Even for more general cases than the FLRW universe, using a vector $\zeta^\mu$ satisfying \eqref{eq:beta},
we can always construct a conserved charge as $S:=E[\zeta]$, which can be interpreted as an entropy in some cases\cite{Aoki:2020nzm},
but whose physical interpretation for a general spacetime is still missing.
It will be interesting and important not only to find a more general physical meaning of $S$ than the entropy but also to 
understand a mechanism working behind this conservation law in general relativity. 

\section{Do gravitational fields carry energies ? }
As a concluding remark, I consider the above question. 
Since our definition of the energy or its generalization  requires a non-zero EMT,
gravitational fields including gravitational waves cannot have such a conserved quantity. 
This seems to contradict a majority's common sense that gravitational waves carry energies.  
After careful consideration, however,
we realize that
there exists no theoretical as well as experimental evidences which unambiguously prove that gravitational waves are carrying energies,\footnote{
It is often said that  an orbital decay for a binary star system is explained by a loss of their energy due to gravitational waves.
More precisely, however, it is confirmed only  that a rate of the orbital decay is consistent with the Einstein equation in some approximation,
where a description of the gravitational wave energy is valid within the approximation.}
as I already discussed 
that a concept of an energy for gravitational fields is not easily justified in general relativity.

Let me be more conservative. Logically it is certainly possible to define another quantity associated with gravitational fields,
which may be regarded as gravitational energy.
Therefore, somebody might find such a new definition of gravitational energy in the future,
which at the same time, is  exchangeable between matter and gravitational fields, so that
only the sum of  matter energy in \eqref{eq:energy} and the gravitational energy is conserved for general spacetime.
Even though such a possibility cannot be ruled out so far,
we have established in our paper\cite{Aoki:2020nzm} that
there {\it always}  exists 
a conserved quantity defined by $E[\zeta]$ with $\zeta$ satisfying \eqref{eq:beta}, 
as carried only by matter.

Let me go back to the start of this essay. The Einstein equation \eqref{eq:EE} clearly tells us a matter $T^{\mu\nu}$ generates a gravitational field.
On the other hand, the gravitational field itself generates no additional gravitational field. 
This structure of the Einstein equation seems naturally to explain why gravitational fields carry no energy. 
 
Finally,  it is my belief that the definition of energy in general relativity is not completely understood yet, as is evidenced by the existence of various proposals for energy in literature. 
It is hoped that the present essay will trigger further analyses of this interesting issue.

\section*{Acknowledgments}
I have enjoyed fruitful collaborations with Drs.~Tetsuya Onogi and Shuichi Yokoyama.
Without them, I can never have written this essay.

\end{document}